\documentclass[runningheads]{llncs}

\usepackage[T1]{fontenc}
\usepackage{siunitx}

\usepackage{booktabs}
\usepackage{amssymb}
\usepackage{graphicx,verbatim}
\usepackage{xcolor}

\begin{document}

\title{MeDi: Metadata-Guided Diffusion Models for Mitigating Biases in Tumor Classification}
\titlerunning{Metadata-Guided Diffusion Models for Histopathology}

\author{%
  David Jacob Drexlin\inst{1,2} \and
  Jonas Dippel\inst{1,3,4} \and
  Julius Hense\inst{1,3} \and
  Niklas Prenißl\inst{5} \and
  Grégoire Montavon\inst{3,6} \and
  Frederick Klauschen\inst{3,5,7,8,9} \and
  Klaus-Robert Müller\inst{1,3,10,11}%
}
\authorrunning{D. Drexlin et al.}

\institute{
Machine Learning Group, Technische Universität Berlin, Germany \and
The Konrad Zuse School of Excellence in Learning and Intelligent Systems (ELIZA), Germany \and 
BIFOLD – Berlin Institute for the Foundations of Learning and Data, Germany \and
Aignostics, Germany \and
Institute of Pathology, Charité – Universitätsmedizin Berlin, Germany \and
Charité -- Universitätsmedizin Berlin, Germany\and
Institute of Pathology, Ludwig-Maximilians-Universität München, Germany \and
German Cancer Research Center (DKFZ) \& German Cancer Consortium (DKTK), Berlin \& Munich Partner Sites \and
Bavarian Cancer Research Center (BZKF), Germany\and
Department of Artificial Intelligence, Korea University, Republic of Korea \and
Max-Planck Institute for Informatics, Germany \\
Correspondence to: \email{\{j.dippel, klaus-robert.mueller\}@tu-berlin.de}}

\maketitle              

\begin{abstract}
Deep learning models have made significant advances in histological prediction tasks in recent years. However, for adaptation in clinical practice, their lack of robustness to varying conditions such as staining, scanner, hospital, and demographics is still a limiting factor: if trained on overrepresented subpopulations, models regularly struggle with 
less frequent patterns, leading to shortcut learning and biased predictions. Large-scale foundation models have not fully eliminated this issue. Therefore, we propose a novel approach explicitly modeling such metadata into a \textbf{Me}tadata-guided generative \textbf{Di}ffusion model framework (MeDi). MeDi allows for a targeted augmentation of underrepresented subpopulations with synthetic data, which balances limited training data and mitigates biases in downstream models. We experimentally show that MeDi generates high-quality histopathology images for unseen subpopulations in TCGA, boosts the overall fidelity of the generated images, and enables improvements in performance for downstream classifiers on datasets with subpopulation shifts. Our work is a proof-of-concept towards better mitigating data biases with generative models.


\end{abstract}

\setcounter{footnote}{0} 

\section{Introduction}
Histopathology is a cornerstone of clinical diagnostics and biomedical research, providing critical insights into disease mechanisms through tissue examination. AI models have shown great potential to advance pathological workflows in recent years \cite{histo-xai-review}. Research studies have shown that deep learning models can predict disease subtypes \cite{campanella2019clinical}, disease progression and outcome \cite{courtiol2019deep}, detect rare cancers \cite{dippel2024ai} and even molecular properties such as mutations \cite{dernbach2024dissecting} and gene expressions \cite{jaume2024hest} to a certain extent. Although models can reach or occasionally even surpass pathologist-grade performance in specific subtasks, they are not widely adopted in the clinical workflow yet \cite{dippel2024ai,histo-xai-review}. The varying conditions across medical centers such as different staining protocols, lab artifacts, scanner type and demographic differences pose unique challenges to machine learning algorithms \cite{histo-xai-review}. If such tissue conditions are correlated with prediction targets, models use them as a shortcut in the training process and fail to generalize \cite{clever-hans,geirhos2020shortcut,howard2021impact,vaidya2024demographic}. Foundation models that are trained with self-supervised learning on large-scale histopathology datasets are proposed as a solution \cite{uni,virchow,rudolfv,atlas}. However, as they not only model the biological features but also these metadata conditions, they can still fail catastrophically when the distribution is highly skewed \cite{kauffmann2024clever,komen2024histopathological,dejong2025currentpathologyfoundationmodels,vaidya2024demographic}. 
Therefore, additional mitigation strategies are needed to prevent such biases and ensure a safe deployment of machine learning models in the clinic.

We propose to explicitly model these conditions by training \textbf{me}tadata-guided generative \textbf{di}ffusion models (MeDi). Instead of conditioning a model only on the disease subclass, we also condition the generation process on readily available metadata attributes, such as the medical center, to be more sensitive to relevant statistical and morphological variations. Furthermore, the MeDi framework provides interpretable controls to balance out the training data distribution during downstream tasks. By contrast, when relying on a single class label, generative models may gravitate toward the ``mode'' of each class distribution, overlooking minority subpatterns/classes and failing to meaningfully increase diversity in the dataset. We demonstrate that (1) MeDi produces higher-quality images compared to class-only conditioning, and (2) can be used to improve generalization capabilities to unseen subpopulations in downstream tasks. The full MeDi pipeline (training, sampling, embedding, and evaluation scripts) is available on GitHub\footnote{\url{https://github.com/David-Drexlin/MeDi}}.

\section{Related Work}
Generative approaches for mitigating domain shifts in medical imaging have received considerable attention in recent years. Early attempts in histopathology often relied on stain normalization \cite{macenko2009method,reinhard2001color} or style-transfer techniques (e.g., CycleGAN-based color normalization) to reduce color variability across laboratories, scanners, and staining protocols \cite{DEBEL2021102004}.

More recent advances in diffusion models have opened up new opportunities for controllable and high-fidelity image synthesis \cite{dhariwal2021diffusion}. For example, Niehues et al.~\cite{kather-diffusion} demonstrated that both fine-tuned Stable Diffusion and GAN-based models can be used to augment datasets in colorectal cancer tasks, achieving improved performance on the CRC-100K dataset. Aversa et al. \cite{aversa2024diffinfinite} extended histological image generation to larger regions with conditioning on class-based segmentation masks, enabling the generation of synthetic whole slide images with pixelwise class annotations.

Beyond class annotations, other conditioning mechanisms in diffusion models have recently been explored. Osorio et al. \cite{osorio2024latent} introduced a latent diffusion framework that leverages cluster-defined morphological features to enhance the fidelity of generated images. Carillo et al. \cite{carrillo2023rna} used bulk RNA sequencing data as a conditioning mechanism. Closest to our own approach, Ktena et al.~\cite{ktena2024generative} demonstrated that diffusion-based augmentation can improve model fairness and robustness in radiology, dermatology, and histopathology. However, their histopathology experiments were limited to breast cancer tissues from the Camelyon17 dataset which offers limited classes (2) and metadata variability (5 hospitals). By contrast, our work spans a broader range of cancer types (32 classes), dozens of medical centers (184), and employs histopathology foundation models in downstream tasks.

\begin{figure}[t]
    \centering
    \includegraphics[width=\linewidth]{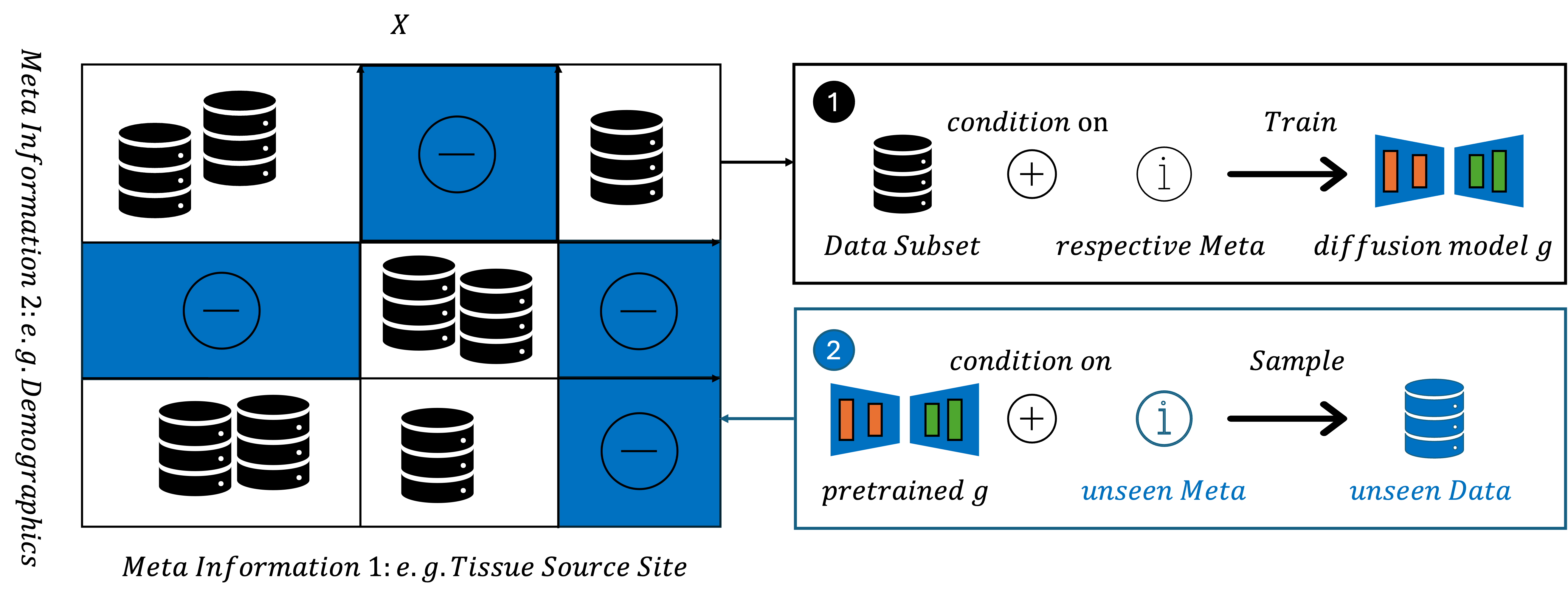}
     \caption{\textbf{MeDi Training and Inference Framework.} During training (1), the model receives real images along with their class labels and metadata. At inference (2), users can condition on arbitrary combinations of class and metadata, enabling the generation of synthetic images for underrepresented subpopulations or unseen subpopulations altogether.}
    \label{fig:model_architecture}
\end{figure}

\section{Method}
In this work, we propose a framework for training metadata-conditioned diffusion models (MeDi) that integrates both class labels (e.g., cancer type) and relevant metadata (e.g., medical center, demographics) into the diffusion process. Conventional class conditioning may capture dominant variations within each class; however, it can fail to model underrepresented or subtle domain shifts. By explicitly modeling additional metadata, MeDi aims to mitigate batch effects in downstream tasks and enhance the diversity of synthetic samples.

Figure~\ref{fig:model_architecture} illustrates the overall framework of MeDi. During training, each image is paired with: a class label (e.g., cancer subtype) and metadata attributes. The model learns the interactions between these attributes and the image features. At inference time, the user can condition the generation process on any valid combination of class and metadata, thereby synthesizing images for underrepresented or unseen subpopulations.

Essentially, our approach can interpolate within the metadata space to balance the distribution of existing meta-class combinations, or extrapolate to produce synthetic samples for combinations absent in the training set. This data-centric approach aims to ''fill in'' distribution gaps and help mitigate biases resulting from skewed datasets. For instance, consider two hospitals: Hospital~A, which submitted lung adenocarcinoma and lung squamous cell carcinoma samples from overwhelmingly Caucasian patients, and Hospital~B, which mostly has samples from African-American patients but exclusively submitted lung squamous cell carcinoma samples. By leveraging metadata-guided diffusion, MeDi can learn to generate realistic synthetic samples for patients with lung adenocarcinoma in Hospital~B, and thus alleviate the imbalance across different unseen subpopulations that are known to bias classifiers \cite{vaidya2024demographic}. By balancing these subpopulations, MeDi helps remove spurious correlations or learnable proxies between metadata and the disease label, reducing Clever Hans effects \cite{clever-hans,kauffmann2024clever} in downstream tasks.

\subsection{Metadata-Conditioned Diffusion UNet}
We build upon a standard 2D UNet architecture from the \emph{diffusers} library \footnote{\url{https://github.com/huggingface/diffusers}}, which progressively denoises an image across multiple residual downsampling and upsampling blocks. 

\paragraph{Metadata Embeddings.} 
Let \(\{\alpha_1, \alpha_2, \dots, \alpha_k\}\) denote the set of metadata attributes associated with each patch. For each \emph{categorical} attribute, we employ a learnable embedding layer that maps each discrete category to a fixed-dimensional vector of dimension \(d_e\). For instance, if the attribute ``medical center''  has \(N_{\text{site}}\) possible values, then each site \(i\) is represented by an embedding vector \(\mathbf{z}_{\text{site}}(i) \in \mathbb{R}^{d_e}\).

\paragraph{Metadata Combination.} 
We concatenate the class label embedding with all metadata embeddings to form a single \emph{conditioning vector}. Let \(\mathbf{z}_{\text{class}} \in \mathbb{R}^{d_{\text{class}}}\) denote the class embedding and let \(\mathbf{z}_{\text{meta},i} \in \mathbb{R}^{d_e}\) denote the embedding of the \(i\)th metadata attribute for \(i=1,\dots,k\). To ensure compatibility with the timestep embedding \(\mathbf{z}_t \in \mathbb{R}^{d_t}\), we require that \(d_{\text{class}} + k\cdot d_e = d_t.\)
The conditioning vector is then defined as

\[
\mathbf{z}_{\text{cond}} = \mathrm{concat}\!\Bigl(\mathbf{z}_{\text{class}},\, \mathbf{z}_{\text{meta},1},\, \dots,\, \mathbf{z}_{\text{meta},k}\Bigr) \in \mathbb{R}^{d_t}.
\]
This vector is added to the timestep embedding:
\[
\mathbf{z}_{\text{final}} = \mathbf{z}_t + \mathbf{z}_{\text{cond}},
\]

which is provided to the UNet’s residual blocks at each down- and upsampling stage. In this way, the diffusion process is conditioned on both the class-specific morphology (e.g., tumor [sub]type) and the domain-specific variations (e.g., scanner artifacts, demographic effects).

\section{Experiments}

\begin{figure}[t]
    \centering
    \includegraphics[width=1\linewidth]{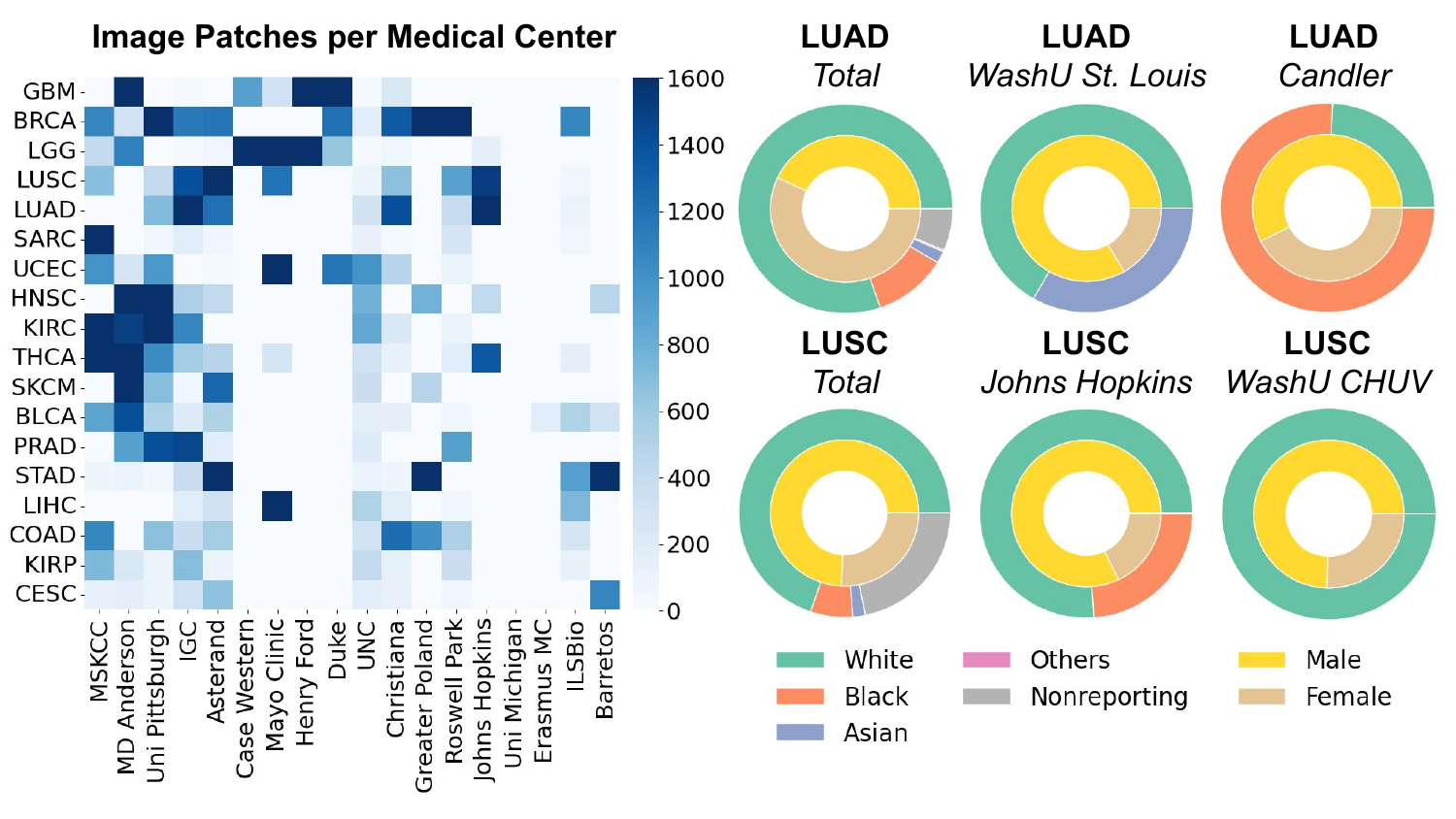}
    \caption{\textbf{Dataset statistics for TCGA-UT}. \textit{Left}: The number of patches for the 18 largest cancer types (vertical axis) and tissue source sites (horizontal axis), capped at 1,600. \textit{Right}: Race and gender distributions for the patches of selected cancer types and tissue source sites. The dataset is highly imbalanced across classes, hospitals, and demographics, with many missing or underrepresented metadata combinations.}
    \label{fig:CLS_TSS_Dist}
\end{figure}

We evaluate our metadata-guided diffusion framework (MeDi) on the TCGA-UT dataset\footnote{zenodo.org/records/5889558, License: Non-Commercial Use: CC-BY-NC-SA 4.0 }~\cite{komura22tcga_uniform}, which provides histopathology tumor patches from 32 distinct cancer types within The Cancer Genome Atlas (TCGA). Each patch is associated with several metadata fields, including \textit{tissue source site (TSS)}, \textit{race}, \textit{gender}, and \textit{age}. In total, the dataset covers 8,726 patient-level records and 271,710 image patches. In terms of metadata dimensions, the dataset contains 626 TSS codes from 184 unique medical centers. This yields a high combinatorial space of $626\times 32 = 20,032$, which is only partially represented. For instance, some cancer types contain fewer than 1,000 patches, while others exceed 20,000; certain TSS are overrepresented, whereas others appear only sporadically as visualized by Figure~\ref{fig:CLS_TSS_Dist}, presenting substantial heterogeneity in both sample sizes and metadata distributions.

\subsection{Dataset Splitting and Training Setup}
\label{subsec:data-split}
To induce an out-of-distribution scenario for the classification tasks,, we design a holdout strategy that excludes specific medical center and demographic combinations from diffusion model training.
We remove certain tissue--demographic configurations from training such that no data leakage can occur through synthetic images. Specifically, for each cancer type (e.g., \texttt{TCGA-CESC}), we randomly select 30\% of the medical center and patient race combinations for the holdout set. 
All patches matching those combinations are excluded from training, leaving 7,061 patient records in the training split and 1,665 in the holdout. 
This ensures that particular site--demographic subpopulations remain completely unseen during model training.

\subsection{Diffusion Model Variants}
We train two distinct diffusion models on the above training set, each for $800,000$ optimization steps at a learning rate of $10^{-4}$ with a batch size of $64$:

\begin{enumerate}
    \item \textbf{Baseline: Class-Only (CLS).}  
    A UNet conditioned solely on the cancer-type label.
    
    \item \textbf{MeDi}: Class + Tissue Source Site (CLS+TSS). 
    A UNet conditioned on the cancer-type label and the medical center (TSS code). Tissue source site is embedded via a learnable layer. 
\end{enumerate}

Both models share the same network capacity and training hyperparameters; they differ only in the presence or absence of the metadata embeddings. We picked the medical center as a proof of concept as it has shown to significantly bias classifiers in past studies \cite{howard2021impact,komen2024histopathological,dejong2025currentpathologyfoundationmodels}.

\subsection{Training Set Fidelity (FID)}
\label{subsec:exp-fid}

\begin{figure}[h!]
    \centering
    \includegraphics[width=.95\linewidth]{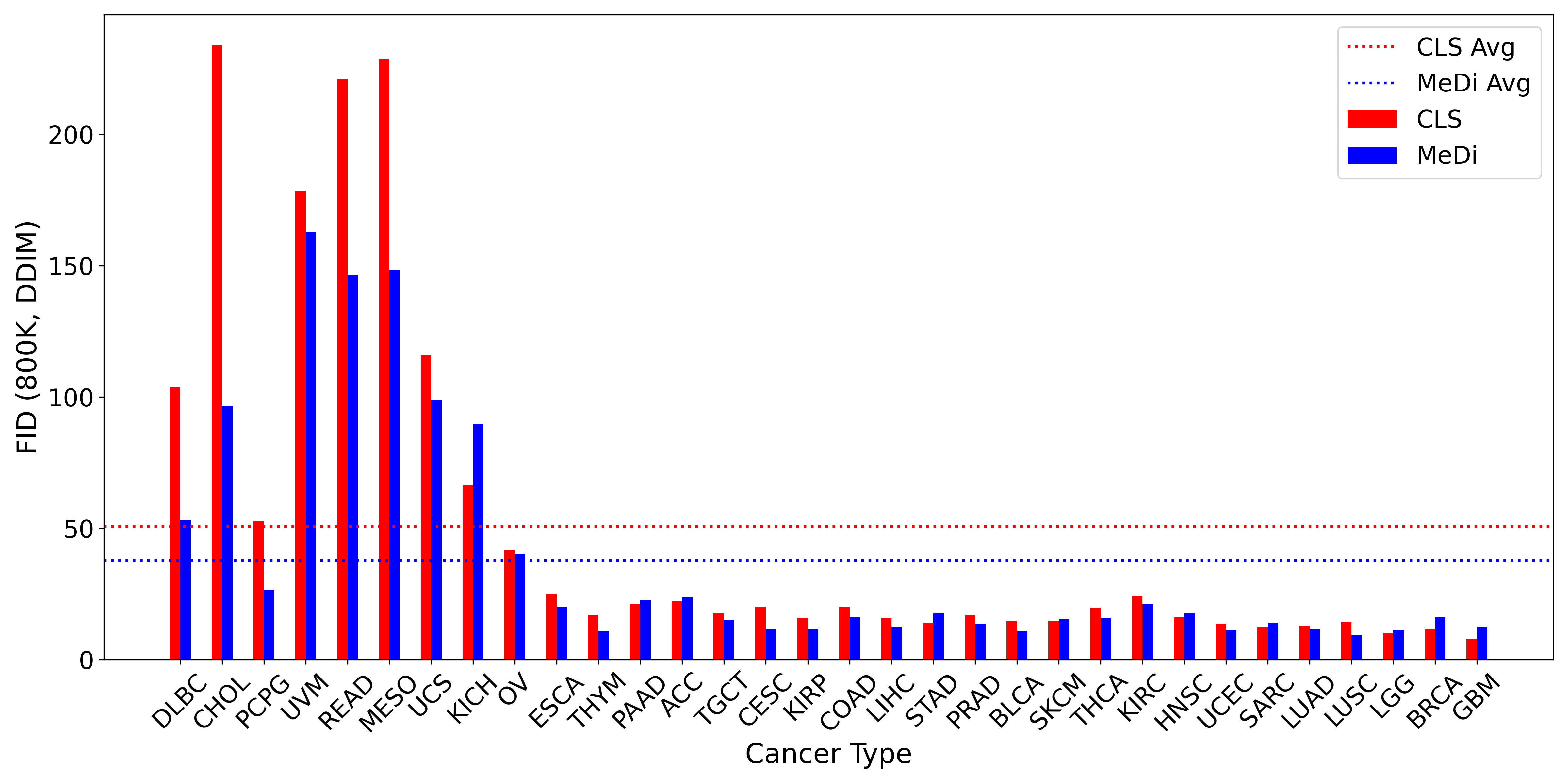}
    \caption{\textbf{FID per cancer type at 800K optimization steps sampled with DDIM}. Depicted are the class-only baseline (CLS, red) and the metadata-conditioned model (MeDi, blue). Cancer types are ordered in descending order based on the number of images in the dataset. Dotted horizontal lines represent the average FID per model: CLS: 50.65, MeDi: 37.73.}
    \label{fig:FID_by_CancerType_BarPlot}
\end{figure}

We first assess how effectively each model reproduces the \emph{training distribution} by measuring the Fréchet Inception Distance (FID) between real and synthetic images. Lower FID scores indicate that synthetic images better match the real data.\\
{\em Setup.}
For each diffusion model, we generate 271,710 synthetic images—matching the exact size and class distribution of the training set - using the DDIM scheduler with 100 inference steps. For each cancer type, we synthesize patches according to the same frequencies of the cancer types (and metadata, when applicable). Thus, both models (CLS, MeDi) produce the same total number of images but differ in how conditioning is applied. Figure~\ref{fig:FID_by_CancerType_BarPlot} breaks down the results per class. 

\begin{itemize}
    \item \textbf{CLS Baseline:} Synthetic patches are generated conditioned solely on the cancer type (CLS).
    \item \textbf{MeDi} Synthetic patches are generated conditioned on both class and the medical center (TSS code in TCGA).
\end{itemize}

\begin{figure}[h!]
    \centering
    \includegraphics[width=.95\linewidth]{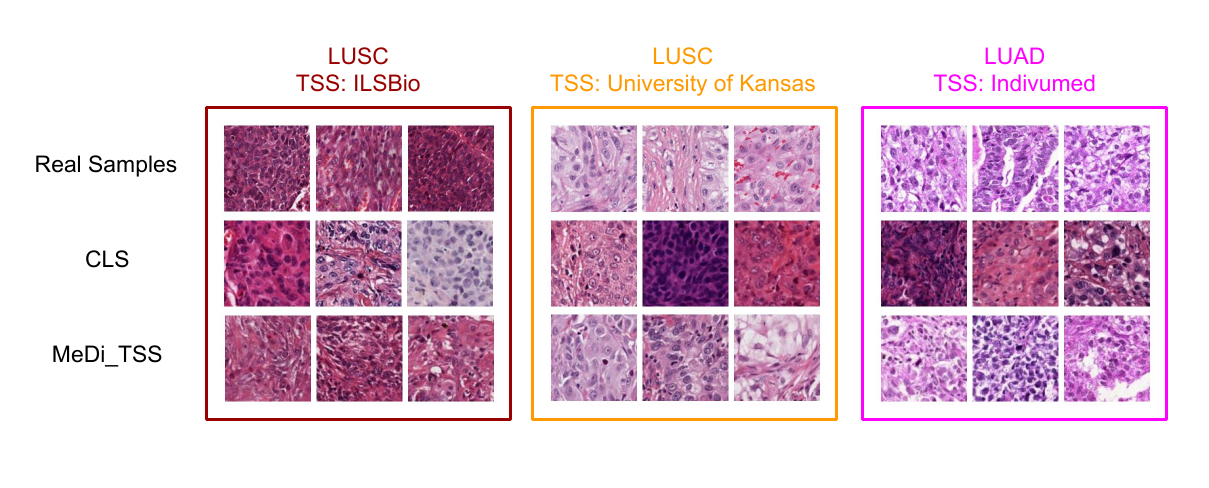}
    \caption{\textbf{LUSC and LUAD samples from different tissue source sites (TSS) along with generated images}. The top row shows real data, while the two other rows show generated samples. MeDi enables capturing the staining mode of the different tissue source sites, while the CLS only model can not be conditioned on a specific TSS and, therefore, is unable to match the real data sample distribution.}
    \label{fig:tss-samples}
\end{figure}

We observe that MeDi achieves a lower average FID than the CLS-only baseline, suggesting that modeling the TSS information helps to produce more faithful synthetic patches. In total, 23 out of 32 cancer types show improved FID scores relative to the baseline, with the most notable improvements occurring in underrepresented cancers that have fewer real samples: metadata conditioning evidently reduces mode collapse by highlighting domain-specific variations and permitting more dataset-specific sampling strategies. This can be seen qualitatively in Figure~\ref{fig:tss-samples}. As the CLS model is unable to generate samples for a specific tissue source site, the MeDi model is able to better capture individual staining patterns of specific subgroups.

\subsection{Performance of Downstream Models on Subpopulation Shift}

Next, we assess whether our generation model can improve downstream task performance by training tumor‐subtype classifiers on three complementary subtyping problems in TCGA‐UT. NSCLC subtyping contrasts lung adenocarcinoma (16,460 patches) with lung squamous cell carcinoma (16,560 patches), both well represented across many centers; RCC subtyping addresses a more imbalanced three‐class setting of clear‐cell (11,650), papillary (6,790), and chromophobe (2,460) renal carcinomas; and uterine cancer classification contrasting infrequent uterine carcinosarcoma (2,120) against abundant endometrial carcinoma (12,480). These tasks span distinct imbalance regimes and metadata–class correlations, thereby testing MeDi’s potential to synthesize patches under realistic subpopulation shifts.

In the training set, each cancer type is sourced from only one distinct TSS. This introduces a correlation between TSS and class label, which mirrors many real-world scenarios where entire data cohorts may come from a single hospital. For both models (CLS and MeDi), we augment the training dataset by adding one synthetic example per real sample. For the CLS model, we generate samples only conditioning on the respective cancer type and sample uniformly over the classes. For MeDi, we generate synthetic samples by conditioning on the class and TSS: we uniformly sample over all possible combinations of selected cancer types and tissue source sites. That is, given real data of cancer type 1 from TSS A and class 2 from TSS B, we generate synthetic samples for (1, A), (1, B), (2, A), and (2, B), filling the missing combinations.
This tailored sampling strategy aims to systematically break the class and metadata correlation. We train a linear classifier on top of the embeddings of $20$ samples per class, extracted with the histopathology foundation model UNI \cite{uni}. The test set contains data from all TSS that the diffusion models have not seen during training, thereby creating a realistic subpopulation shift scenario. We repeat the model training for each possible combination of TSS from the training set as defined in Section \ref{subsec:data-split}.

\begin{table}[h!]
    \centering
    \caption{Linear probing results averaged over all possible tissue source site combinations in the training set. \emph{Overall} indicates total balanced accuracy over the entire test set while for \emph{TSS AVG}, the accuracy is computed for each TSS and averaged. The depicted variance shows standard error and best results are in bold.}
    \label{tab:luad_lusc}
    \scriptsize
    \begin{tabular}{|l|cc|cc|cc|}
        \toprule &  \multicolumn{2}{|c|}{NSCLC} &  \multicolumn{2}{|c|}{RCC} & \multicolumn{2}{|c|}{Uterine} \\
        Method & Overall & TSS AVG & Overall & TSS AVG & Overall & TSS AVG\\
        \midrule
No syn. data & 74.31 $\pm$ \tiny{0.46} & 75.76 $\pm$ \tiny{0.51} & 73.25 $\pm$ \tiny{0.32} & 81.74 $\pm$ \tiny{0.24} & 67.73 $\pm$ \tiny{1.11} & 62.82 $\pm$ \tiny{2.24} \\
CLS only & 79.99 $\pm$ \tiny{0.31} & 81.72 $\pm$ \tiny{0.33} & \textbf{78.21 $\pm$ \tiny{0.27}} & \textbf{84.64 $\pm$ \tiny{0.14}} & 67.77 $\pm$ \tiny{0.89} & \textbf{70.80 $\pm$ \tiny{1.37}} \\
MeDi & \textbf{81.37 $\pm$ \tiny{0.27}} & \textbf{83.04 $\pm$ \tiny{0.30}} & 74.87 $\pm$ \tiny{0.24} & \textbf{84.17 $\pm$ \tiny{0.14}} & \textbf{74.86 $\pm$ \tiny{0.84}} & \textbf{71.21 $\pm$ \tiny{1.58}} \\
\bottomrule
    \end{tabular}
\end{table}
Table \ref{tab:luad_lusc} shows the average balanced accuracy and the tissue source site (TSS)–averaged accuracy across all experimental runs. We find that synthetic data augmentation significantly improves performance to unseen tissue source sites in low data regimes. MeDi can further improve accuracy for NCLSC and Uterine classification while being roughly on par for RCC subtyping when considering the average TSS performance. This showcases that targeted data generation with MeDi can lead to better classifiers. In future work, we want to investigate if similar results can be achieved on larger datasets and in more general settings.
\section{Conclusion}
Machine Learning solutions in the medical domain suffer from a lack of robustness to varying conditions (e.g.~staining, scanner, hospital demographics), specifically  over and underrepresented patient subpopulations challenge training success. 
This work shows that explicitly conditioning diffusion models on relevant metadata—such as hospital site yields two key benefits in computational histopathology: (1) improved fidelity of synthetic images to real-world distributions, and (2) that generative data augmentation can complement discriminative foundation models by selectively balancing subpopulation shifts.
\noindent In future work, we will explore further metadata factors (e.g., scanner type, patient age, gender, and race), larger-scale training regimes, and classifier-free guidance to further assess and improve both the realism and utility of synthetic histopathology data.

\section*{Acknowledgements}
This work was in part supported by the German Ministry for Education and Research (BMBF) under Grants 01IS14013A-E, 01GQ1115, 01GQ0850, 01IS18025A, 031L0207D, and 01IS18037A. K.R.M. was partly supported by the Institute of Information \& Communications Technology Planning \& Evaluation (IITP) grants funded by the Korea government (MSIT) (No. 2019-0-00079, Artificial
Intelligence Graduate School Program, Korea University and No. 2022-0-00984, Development of Artificial Intelligence Technology for Personalized Plug-and-Play Explanation and Verification of Explanation). D.J.D. is supported by the Konrad Zuse School of Excellence in Learning and Intelligent Systems (ELIZA) through the DAAD programme Konrad Zuse Schools of Excellence in Artificial Intelligence, sponsored by the Federal Ministry of Education and Research.

The results shown here are in whole or part based
upon data generated by the TCGA Research Network:
https://www.cancer.gov/tcga.

\bibliographystyle{splncs04}
\bibliography{references}

\end{document}